\documentclass[runningheads]{svmult}

\usepackage{makeidx}   
\usepackage{graphicx}  
\usepackage{subeqnar}  
\usepackage{multicol}  
\usepackage{physprbb}  
\makeindex             

\begin{document}
\title*{Proposed nomenclature for Extragalactic Planetary Nebulae}
\toctitle{Proposes nomenclature for
\protect\newline Extragalactic Planetary Nebulae}
%
%
\titlerunning{Extragalactic PN nomenclature}
%
\author{Quentin A Parker\inst{1,2}
\and Agnes Acker\inst{2}}
%
\authorrunning{Parker \& Acker}
%
%
\institute{Macquarie University/Anglo-Australian Observatory, Sydney, Australia,
qap@ics.mq.edu.au
\and Observatoire de Strasbourg, France, acker@astro-u-strasbg.fr}

\maketitle              

\section{Introduction}

The ability to identify and distinguish between the wide variety of
celestial objects benefits from 
application of a systematic and logical nomenclature.
This often includes value-added information within the naming convention which can aid in 
placing the object positionally –either via an RA/DEC or l,b concatenation.
All new nomenclatures should be created following IAU guidelines. 
However as the number density of specific object types
on the sky increases, as in the case of PN in external galaxies, 
a useful positional identifier becomes problematic.
This brief but timely  paper attempts to progress the debate on this vexing issue 
for the case of extragalactic planetary nebulae
(PNe). There is a clear need to rationalise the current ad-hoc system now that many
thousands of Extragalactic PN are being discovered.

\section{Existing structures and conventions}

A comprehensive web site exists at the Centre de Donn\'{e}e astronomiques de Strasbourg (CDS) 
to assist the process… of astronomical nomenclature and a
registration form is available at http://www.vizier.u-strasbg.fr/viz-bin/DicForm…. 
The current convention for galactic PN is to use `PN Glll.l+/-bb.b'.
However, the positional information included is no longer sufficient to discriminate amongst all the 
known galactic PNe with the advent of MASH (Macquarie/AAO/Strasbourg H-alpha Planetary Nebula 
project) which has added $\sim1000$ new Galactic PNe  including several PNe `pairs' 
within an arcminute or so of each other (Parker et al. 2003).
PN Glll.l+/-bb.b a,b,c etc is used in the `unlikely''
event of PNe being found in such close proximity….
This untidy fix demonstrates the need to try to future proof such 
designations against use of too crude positional 
identifiers if they are to be used to any effect in the designation at all.

\subsection{Application of current convention}
A first pass proposal has been suggested: PN X JHHMMSSs+DDMMSS. This includes 
a PN prefix `PN' followed by  a galaxy identifier
`X' (e.g. NGC7793) then `J' to indicate equinox J2000 and then an RA/DEC positional identifier. 
Let''s look at the initial proposal applied to a PN in galaxy NGC7793:\\
We would have: PN NGC7793 J2357499-323520.
This is quite a mouthful even if we replace NGC with N. 
Thousands of PNe are being found in galaxies at greater 
distance which are of small angular extent. An individual PN would require 
an extremely long indentifier if positional concatenation 
is to be used to any effective purpose.
Hence PN NGC7793 J23574998-3235206 at least may be required. Furthermore,
any identifier should be future proofed to work
when PNe are found a hundredth of an arcsecond from another in a distant galaxy…. 
Perhaps an RA/DEC concatenation is not the best way forward.

\section{What's in a name?}
The unwieldly nature of a simple logical extension of the current system and its potential
ambiguity has motivated us to investigate a simple
alternative. Clearly the host galaxy is a key parameter 
immediately placing the PNe in context.
Then we need to identify the object as an extragalactic PN (or PN candidate) and
we need to discriminate it from all the other (perhaps hundreds or even thousands) of PN 
that may be discovered in an individual galaxy or its environs.

We propose a simple `EPN' prefix to denote extragalactic PN, 
then a galaxy identifier and then a running number, perhaps in order of PN discovery epoch.
The greater the number the more recent the discovery.
Also the largest number immediately tells you approximately 
how many PN have been found in the particular galaxy to date.
The name is also of manageable length and
can be easily cross-referenced to all relevant data and accurate 
positions which may be updated as astrometry improves. 
Crucially the ID will remain the same.

\subsection{Proposed new nomenclature}
EPN-NGC7793-1, EPN-N7793-1157  or EPN-IC5175-32 etc and that''s it!
The key `EPN'' prefix can be used to search entire CDS databases and will naturally pick up
all extragalactic PN regardless of galaxy affiliation. 
We assume that any galaxy will always have a suitable identifier and that 
for brevity NGC can be concatenated to `N'. 
The hyphens serve as useful delimiters between the key nomenclature
components and assist in readability. 

\section{Problems and issues}
We have successfully registered the EPN acronym with the relevant IAU commission but
there are of course some problems with this simple system and no
scheme will ever satisfy the whole community.
However there is strong community support for the
essence of the new proposal together with a keen desire to urgently sort out the naming 
convention. Any adopted system cannot function by self-regulation 
so each group/discovery team
working on extragalactic PN can present their own identifiers for each object as their
`usual names' and will not be expected to assign any adopted 
standardised nomenclature themselves.
We propose that a small committee be tasked to assign nomenclatures 
to new extragalactic PNe. This would convene once
or twice a year to incorporate all results published in the interim. 
An electronic only catalogue of such EPN will be
made available. The preliminary committee membership is suggested as Acker, Jacoby and Parker.
 
\subsection{Intracluster PN}
How do we name them?
The most supported suggestion is to
attach them to the closest named galaxy. 
Many felt we can simply retain EPN-N7793-2415 for example. This is especially 
germane as the question of where does a galaxy end and the intergalactic medium begin? 
An alternative is to simply
divide the sky into a few hundred cells and allocate PN accordingly. 
This was seen less favourably. PN associated with free-floating globular clusters or 
whatever would still work IF we assume such objects will have a name.

\subsection{Contaminants}
In many extragalactic PN projects many `new' PN are candidates, 
being selected on the basis of a single line (usually [OIII] 5007\AA).
When such candidates are subsequently identified as 
contaminants in a numbered scheme `1-n' a
removed object `m'' will leave a gap in what may have been a nice ordered sequence.
This does' not matter. Even in commercial data-bases such gaps are quite acceptable and common 
and their number can be easily determined.
Furthermore we need only accept a PN and assign it the new nomenclature once it has been confirmed.
Also it does not matter if  pet designations for specific projects are also retained, 
after all any SIMBAD query for a Galactic PN can turn up a plethora of individual 
identifiers for the same object.

\section{Conclusions}
There is an urgent need to adopt a working extragalactic PN nomenclature that
can then be implemented and adhered to. An extension to existing conventions could easily yield an 
identifier containing
nearly 30 characters. This was felt by most to be too unwieldly and anonymous for general use. 
We propose a simple alternative whilst not perfect is tractable and easy to understand. It will 
require careful management. We commend it to the community. 

\noindent{\bf Acknowledgements}
The authors are grateful for the positive feedback from the community and the useful comments on an 
earliler version of this paper by George Jacoby.

%

\end{document}